\documentclass[prd,twocolumn,showpacs,amsmath,amssymb,nofootinbib]{revtex4-2}

\usepackage[utf8]{inputenc}
\usepackage{booktabs}%my own
\usepackage{feynmp}
\usepackage{epsf}
\usepackage{graphicx}
\usepackage{epsfig}
\usepackage{xcolor}
\usepackage{subfigure}
\usepackage{pstricks}
\usepackage{pst-node}
\usepackage{rotating}
\usepackage{graphics}
\usepackage{latexsym}
\usepackage{dcolumn}
\usepackage{bm}
\usepackage{times}
\usepackage{indentfirst}
\usepackage{float}
\usepackage{overpic}
\usepackage{amsmath}
\usepackage{appendix}
\usepackage{upgreek}
\usepackage{chngcntr}
\usepackage{amssymb}%
\usepackage{pifont}%
\usepackage[top=0.5in,bottom=0.8in,left=0.9in,right=0.9in]{geometry}
\usepackage{subfigure}
\usepackage{lineno}

\usepackage{multirow}
\usepackage{makecell}

\usepackage{hyperref}
\hypersetup{
	colorlinks=true,
	linkcolor=blue,
	filecolor=blue,
	urlcolor=blue,
	citecolor=blue,
}

\newcommand{\BR}{\mathcal{B}}

\newcommand{\jpsi}{J/\psi}
\newcommand{\etac}{\eta_c}
\newcommand{\kk}{K^+K^-}
\newcommand{\pp}{\pi^+\pi^-}
\newcommand{\ppp}{\pi^+\pi^-\pi^0}
\newcommand{\pip}{\pi^+}
\newcommand{\pim}{\pi^-}
\newcommand{\piz}{\pi^0}

\newcommand{\Mkkppp}{M_{K^{+}K^{-}\ppp}}
\newcommand{\Mppp}{M_{\pi^+\pi^-\pi^0}}
\newcommand{\Mkk}{M_{K^+K^-}}
\newcommand{\dedx}{{\rm d}E/{\rm d}x}
\newcommand{\costhe}{\cos\theta}
\def\TeV{\ifmmode {\mathrm{\ Te\kern -0.1em V}}\else
                   \textrm{Te\kern -0.1em V}\fi}%
\def\GeV{\ifmmode {\mathrm{\ Ge\kern -0.1em V}}\else
                   \textrm{Ge\kern -0.1em V}\fi}%
\def\MeV{\ifmmode {\mathrm{\ Me\kern -0.1em V}}\else
                   \textrm{Me\kern -0.1em V}\fi}%
\def\keV{\ifmmode {\mathrm{\ ke\kern -0.1em V}}\else
                   \textrm{ke\kern -0.1em V}\fi}%
\def\eV{\ifmmode  {\mathrm{\ e\kern -0.1em V}}\else
                   \textrm{e\kern -0.1em V}\fi}%

\let\gev=\GeV
\let\mev=\MeV

\def\TeVc{\ifmmode {\mathrm{\ Te\kern -0.1em V}/c}\else
                   {\textrm{Te\kern -0.1em V}/$c$}\fi}%
\def\GeVc{\ifmmode {\mathrm{\ Ge\kern -0.1em V}/c}\else
                   {\textrm{Ge\kern -0.1em V}/$c$}\fi}%
\def\MeVc{\ifmmode {\mathrm{\ Me\kern -0.1em V}/c}\else
                   {\textrm{Me\kern -0.1em V}/$c$}\fi}%
\def\keVc{\ifmmode {\mathrm{\ ke\kern -0.1em V}/c}\else
                   {\textrm{ke\kern -0.1em V}/$c$}\fi}%
\def\eVc{\ifmmode  {\mathrm{\ e\kern -0.1em V}/c}\else
                   {\textrm{e\kern -0.1em V}/$c$}\fi}%

\let\gevc=\GeVc

\def\TeVcc{\ifmmode {\mathrm{\ Te\kern -0.1em V}/c^2}\else
                   {\textrm{Te\kern -0.1em V}/$c^2$}\fi}%
\def\GeVcc{\ifmmode {\mathrm{\ Ge\kern -0.1em V}/c^2}\else
                   {\textrm{Ge\kern -0.1em V}/$c^2$}\fi}%
\def\MeVcc{\ifmmode {\mathrm{\ Me\kern -0.1em V}/c^2}\else
                   {\textrm{Me\kern -0.1em V}/$c^2$}\fi}%
\def\keVcc{\ifmmode {\mathrm{\ ke\kern -0.1em V}/c^2}\else
                   {\textrm{ke\kern -0.1em V}/$c^2$}\fi}%
\def\eVcc{\ifmmode  {\mathrm{\ e\kern -0.1em V}/c^2}\else
                   {\textrm{e\kern -0.1em V}/$c^2$}\fi}%

\let\gevcc=\GeVcc
\let\mevcc=\MeVcc

\def\cm{\ifmmode  {\mathrm{\ cm}}\else
                   \textrm{~cm}\fi}%

		\newcommand{\bfg}{\begin{figure}}
			\newcommand{\efg}{\end{figure}}
		\newcommand{\bitm}{\begin{itemize}}
			\newcommand{\eitm}{\end{itemize}}
		\newcommand{\bnum}{\begin{enumerate}}
			\newcommand{\enum}{\end{enumerate}}
		\newcommand{\btbl}{\begin{table}}
			\newcommand{\etbl}{\end{table}}
		\newcommand{\btbu}{\begin{tabular}}
			\newcommand{\etbu}{\end{tabular}}
		\newcommand{\bcl}{\begin{center}}
			\newcommand{\ecl}{\end{center}}
		
		\newcommand{\beq}{\begin{equation}}
			\newcommand{\eeq}{\end{equation}}
		\newcommand{\beqr}{\begin{eqnarray}}
			\newcommand{\eeqr}{\end{eqnarray}}
\begin{document}
	
%\title{\boldmath Search for the decay $J/\psi \to \gamma D^0/\bar{D}^0$}
\title{\boldmath Evidence of doubly OZI-suppressed decay $\etac \to \omega\phi$ in the radiative decay $\jpsi \to \gamma\etac$}
%\title{\boldmath Branching Fraction Measurement of the $\eta_{c} \to \omega\phi$ in $J/\psi \to \gamma\eta_{c}$ decay}

\author{
%% Saved at => 2024-11-05
M.~Ablikim$^{1}$, M.~N.~Achasov$^{4,c}$, P.~Adlarson$^{76}$, X.~C.~Ai$^{81}$, R.~Aliberti$^{35}$, A.~Amoroso$^{75A,75C}$, Q.~An$^{72,58,a}$, Y.~Bai$^{57}$, O.~Bakina$^{36}$, Y.~Ban$^{46,h}$
% ...
\\
\vspace{0.2cm}
(BESIII Collaboration)\\
\vspace{0.2cm} {\it
$^{1}$ Institute of High Energy Physics, Beijing 100049, People's Republic of China\\
% ...
$^{4}$ Budker Institute of Nuclear Physics SB RAS (BINP), Novosibirsk 630090, Russia\\
% ...
$^{35}$ Johannes Gutenberg University of Mainz, Johann-Joachim-Becher-Weg 45, D-55099 Mainz, Germany\\
$^{36}$ Joint Institute for Nuclear Research, 141980 Dubna, Moscow region, Russia\\
% ...
$^{46}$ Peking University, Beijing 100871, People's Republic of China\\
% ...
$^{57}$ Southeast University, Nanjing 211100, People's Republic of China\\
$^{58}$ State Key Laboratory of Particle Detection and Electronics, Beijing 100049, Hefei 230026, People's Republic of China\\
% ...
$^{72}$ University of Science and Technology of China, Hefei 230026, People's Republic of China\\
% ...
$^{75}$ University of Turin and INFN, (A)University of Turin, I-10125, Turin, Italy; (B)University of Eastern Piedmont, I-15121, Alessandria, Italy; (C)INFN, I-10125, Turin, Italy\\
$^{76}$ Uppsala University, Box 516, SE-75120 Uppsala, Sweden\\
% ...
$^{81}$ Zhengzhou University, Zhengzhou 450001, People's Republic of China\\
\vspace{0.2cm}
$^{a}$ Deceased\\
% ...
$^{c}$ Also at the Novosibirsk State University, Novosibirsk, 630090, Russia\\
% ...
$^{h}$ Also at State Key Laboratory of Nuclear Physics and Technology, Peking University, Beijing 100871, People's Republic of China\\
% ...
%% ends here %%
}
}

\date{\today}

\renewcommand{\abstractname}{}
\begin{abstract}
	Using a sample of $(10087\pm44) \times 10^{6}$  $\jpsi$ events collected with the BESIII detector at the BEPCII collider, 
the first evidence for the doubly OZI-suppressed decay $\eta_{casdasd} \to \omega\phi$ is reported with a significance of 4.0$\sigma$.
The branching fraction of $\etac \to \omega\phi$ is measured to be $\BR(\etac \to \omega\phi) = (3.86 \pm 0.92 \pm 0.62) \times 10^{-5}$,
where the first uncertainty is statistical and
the second is systematic.
This result provides valuable insights into the underlying mechanisms of charmonium decays, particularly for processes such as 
$\etac \to VV$ (where $V$ represents a vector meson). 

\end{abstract}

\maketitle

\section{INTRODUCTION}
\label{sec: intro}

The $\etac$, as the lightest and $S$-wave spin-singlet charmonium state, plays a fundamental role in the study of charmonium decays. 
Although $\eta_c$ was observed by the Crystal Ball experiment~\cite{Partridge1980} more than forty years ago, 
its properties remain under investigation.
%was first observed by Crystal Ball in 1980~\cite{Partridge1980}, the properties of this lowest S-wave spin singlet charmonium state is still under investigation.
%Until now its exclusively measured decay modes sum up to about 60\% only,
To date, only the exclusive decay modes with a total branching fraction of approximately 60\% have been experimentally observed,
many of which are measured with limited precision and large uncertainties~\cite{pdg}.
As a result, our understanding of the $\etac$ properties remains incomplete.

The decays of $\etac$ into vector meson pairs ($\etac \to VV$, where $V$ denotes a vector meson), which are
 Okubo-Zweig-Iizuka (OZI) suppressed processes~\cite{Okubo:1963fa}, have long posed a significant puzzle in charmonium physics. 
According to the Helicity Selection Rule (HSR)~\cite{Brodsky1981}, these decays are highly suppressed at leading order in Quantum Chromodynamics (QCD).
The branching fractions of $\etac \to \omega\omega$ and $\etac \to \phi\phi$ have been well measured by 
the Beijing Spectrometer III (BESIII) experiment~\cite{Ablikim2019,Ablikim2017a} and Belle experiment~\cite{PhysRevLett.108.232001}.
However, both results are significantly larger than theoretical predictions~\cite{Sun2011}, challenging explanations based solely on the quark model and perturbative QCD calculations. 
Non-perturbative models, such as the intermediate meson exchange model~\cite{Zhao2006}
and the charmonium light Fock component admixture model~\cite{Feldmann2000}, have been proposed as potential solutions. Nevertheless, further investigations are required to fully understand these processes.

\begin{figure}[htp]
  \begin{center}
    \subfigure{\includegraphics[width=0.7\linewidth]{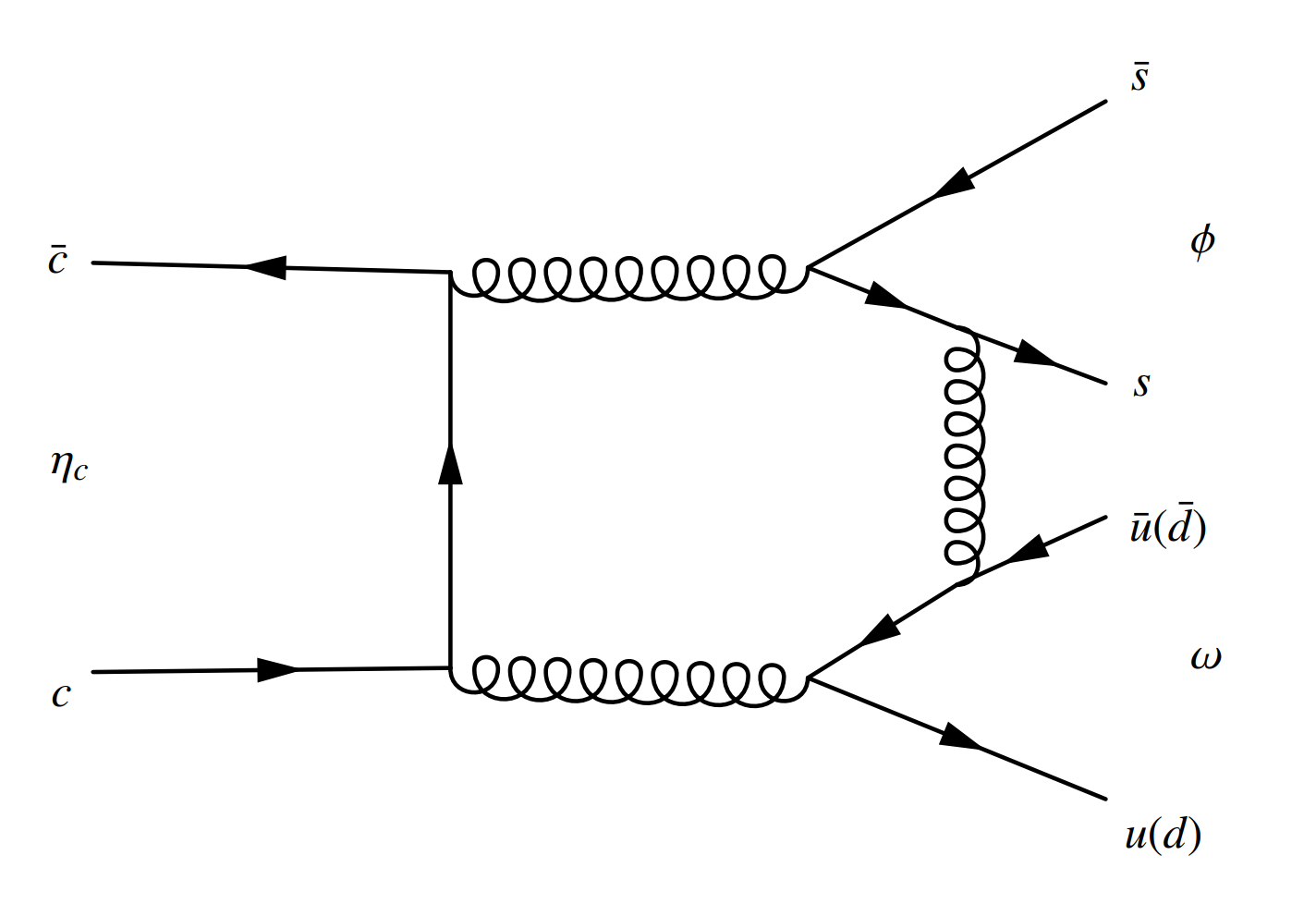}}
    \caption{
			An example of Feynman diagram for $\etac \to \omega\phi$.
      }
    \label{fig:feynman_etactoop}
  \end{center}
\end{figure}

The decay $\etac \to \omega\phi$, which proceeds via a doubly OZI (DOZI) suppressed process as shown by the Feynman diagram in Fig.~\ref{fig:feynman_etactoop}, is expected to provide deeper insights into the underlying mechanisms of decays $\etac \to VV$. Such insights could offer valuable constraints for theoretical models. Several well-established models have been developed to study the mechanisms of DOZI decays in $\jpsi$ and $\etac$ systems~\cite{Seiden1988,Haber1985}.
However, among the DOZI decays of $\etac$, the decay $\etac \to \omega\phi$ has neither been observed experimentally nor investigated theoretically. 
To date, only an upper limit on its branching fraction has been reported by BESIII, based on a sample of  $(223.7\pm 1.4)\times 10^6$ $J/\psi$ events~\cite{Ablikim2017a}.

%Thus, we hope this research will provide parameter input for the corresponding establishment of theoretical model, and contribute to a deeper understanding of $\eta_c \to VV$.

%Though $\eta_c \to VV$ decay modes are mostly Okubo-Zweig-Iizuka (OZI) suppressed,
%Singly OZI (SOZI) suppressed decays $\eta_c \to \omega\omega$ and $\eta_c \to \phi\phi$ have already been well measured by BESIII~\cite{Ablikim2019,Ablikim2017a}.
%However, $\eta_c \to \omega\phi$ decay is doubly OZI (DOZI) suppressed as shown by the Feynman Diagram in Figure~\ref{fig:feynman_etactoop}.

%As a result, only an upper limit has been given by BESIII so far~\cite{Ablikim2017a}.

%The former research of $\eta_c \to \omega\phi$ decay was subjected to the low statistic.
In this paper, utilizing a significantly larger data sample of $(10087\pm 44) \times 10^6$ $\jpsi$ events~\cite{Ablikim2022} accumulated at the BESIII detector,
we present the first evidence for $\etac \to \omega\phi$ and determine its branching fraction, benefiting from both the increased data sample and an improved method for background suppression.

%It's an good opportunity for us to search for the decay of $\eta_c \to \omega\phi$.
\section{BESIII DETECTOR AND MONTE CARLO SIMULATION}
\label{sec:detmc}
The BESIII detector~\cite{Ablikim:2009aa} records symmetric $e^+e^-$ collision events
provided by the BEPCII storage ring~\cite{Yu:IPAC2016-TUYA01}
in the center-of-mass energy ranging from 1.84 to 4.95~GeV,
with an achieved peak luminosity of $1.1 \times 10^{33}\;\text{cm}^{-2}\text{s}^{-1}$  at $\sqrt{s} = 3.773\gev$.
BESIII has collected large data samples in this energy region~\cite{Ablikim:2019hff}. The cylindrical core of the BESIII detector covers 93\% of the full solid angle and consists of a helium-based multilayer drift chamber~(MDC), a plastic scintillator time-of-flight system~(TOF), and a CsI(Tl) electromagnetic calorimeter~(EMC),
which are all enclosed in a superconducting solenoidal magnet providing a 1.0~T magnetic field. 
The magnetic field was 0.9~T in 2012, which affects 11\% of the total $\jpsi$ data.
The solenoid is supported by an
octagonal flux-return yoke with resistive plate counter muon
identification modules interleaved with steel.
%The acceptance of charged particles and photons is 93\% over $4\pi$ solid angle.
The momentum resolution of charged particle at $1\gevc$ is
$0.5\%$, and the  $\dedx$ resolution is $6\%$ for electrons from Bhabha scattering.
The EMC measures photon energies with a resolution of $2.5\%$ ($5\%$) at $1\gev$ in the barrel (end cap)
region. The time resolution in the TOF barrel region is 68~ps, while
that in the end cap region was 110~ps. The end cap TOF
system was upgraded in 2015 using multigap resistive plate chamber
technology, provides a time resolution of 60~ps, which benefits 87\% of the data used in this analysis~\cite{tof1,tof2,tof3}.

Simulated data samples produced with a {\sc geant4}-based~\cite{geant4} Monte Carlo (MC) package,
which includes the geometric description of the BESIII detector and the
detector response, are used to optimize the selection criteria, to determine the detection efficiencies
and to estimate backgrounds. In the simulation, the beam
energy spread and initial state radiation in the $e^+e^-$
annihilations are modeled with the generator {\sc kkmc}~\cite{ref:kkmc}.
An inclusive MC sample consisted of $10^{10}$ $\jpsi$ events, including the production of the $\jpsi$ resonance, are generated to study the backgrounds.
All particle decays are modelled either with {\sc evtgen}~\cite{ref:evtgenl, ref:evtgenp} using branching fractions 
taken from the Particle Data Group (PDG)~\cite{pdg}, when available,
or otherwise estimated with {\sc lundcharm}~\cite{ref:lundcharm1, ref:lundcharm2} package.
%{\it [ORIGINAL:
%The known decay modes are modelled with {\sc
%evtgen}~\cite{ref:evtgen} using BFs taken from the
%Particle Data Group~\cite{pdg}, and the remaining unknown charmonium decays
%are modelled with {\sc lundcharm}~\cite{ref:lundcharm}.] }
Final state radiation from charged final state particles is incorporated using the {\sc photos} package~\cite{photos}.
%To estimate the detection efficiency and optimize the event selection criteria, 
The signal events from $\jpsi \to \gamma \etac$ are generated with 
the modified JPE model~\cite{Mitchell2009,Brambilla2006}.
In this model, the M1 transition is assumed to follow the form factor $E^{3}_{\gamma} \  \times$ exp$(-E^{2}_{\gamma}/8\beta^{2})$ with $\beta = 65.0\pm2.5\mev$.
Here, $E_{\gamma}$ represents the energy of the radiative photon.
These parameters also follow the previous results~\cite{Mitchell2009} with the $\etac$ mass  2984.1$\mevcc$ and the width 30.5$\mev$.
%means that the $M_{\eta_{c}}$ is 2982.2 MeV/$c^{2}$ and $\Gamma_{\eta_{c}}$ is 31.5 MeV/$c^{2}$.
The decays $ \etac \to \omega\phi $, $\omega \to \ppp, \piz \to \gamma\gamma$ and $\phi \to \kk$ are generated with the {\sc evtgen} generator ~\cite{ref:evtgenl, ref:evtgenp} model.
Specifically, the $ \etac \to \omega\phi $ decay is generated with HELAMP model; 
the $\omega\to\ppp$ decay is generated by the OMEGA DALITZ model~\cite{PhysRevD.98.112007}; 
and the $\phi \to \kk$ decay is generated by the VSS model.

%The decay $ \eta_{c} \to \omega\phi $ is generated with HELAMP event generator with parameters set to be \{1.0, 0.0, 0.0, 0.0, -1.0, 0.0\}.
%For $\omega \to \pi^{+}\pi^{-}\pi^{0}$ decay mode, the OMEGA\_DALITZ model is applied,
%and for $\phi \to K^{+}K^{-}$ decay mode, the VSS model is applied.

\section{EVENT SELECTION}
\label{EvtSel}

\subsection{Common selection}
\label{sub: comsel}

In this analysis, the $\eta_c$ candidates are produced via the radiative decay $J/\psi \to \gamma\eta_c$.
%The $\etac$ candidates are produced by $\jpsi$ radiative transition. 
Consequently, the $\etac \to\omega \phi$ candidates are selected via the final state of $3\gamma\pip\pim K^{+}K^{-}$. Candidate events are required to have exactly four charged tracks with zero net charge and at least three photon candidates.
%In this paper, the decay $\jpsi \to \gamma \etac$, $\etac \to \omega\phi$ is reconstructed by detecting a radiative photon, the  decay $\omega \to \ppp$ and $\phi \to \kk$.
%The candidate events are selected by requiring exactly one pair of $\pp$, one pair of $\kk$, 

Charged tracks detected in the MDC are required to satisfy $|\costhe|<0.93$, where $\theta$ is the polar angle defined with respect to the $z$-axis of the BESIII detector.
%which is the symmetry axis of the MDC.
The distance of closest approach to the interaction point (IP) must be less than 10\,cm along the $z$-axis, and less than 1\,cm in the transverse plane.
Particle identification~(PID) for charged tracks combines the specific ionization energy loss~(d$E$/d$x$) measured by the MDC and the flight time in the TOF to form the likelihood values $\mathcal{L}(h)$ for  $h=K$ and $\pi$ hypotheses, individually.
Charged kaons and pions are identified by requiring $\mathcal{L}(K)>\mathcal{L}(\pi)$ and $\mathcal{L}(\pi)>\mathcal{L}(K)$, respectively.
%After PID, a primary vertex fit for the selected $K^{+}K^{-}\pip\pim$ is performed, and a successful vertex fit is required.

Photon candidates are identified using isolated showers in the EMC.
The deposited energy of each shower must be more than 25$\mev$ in the barrel region ($|\costhe|< 0.80$) or more than 50$\mev$ in the end cap region ($0.86 <|\costhe|< 0.92$).
To exclude showers originating from charged tracks,
the opening angle subtended by the EMC shower and the position of the closest charged track at the EMC must be greater than 10 degrees as measured from the IP.
To suppress electronic noise and showers unrelated to the event, the difference between the EMC time and the event start time is required to be within [0, 700]\,ns.

A primary vertex fit is performed for the selected $\kk\pp$ candidate, and a successful vertex fit is required.
To improve the resolution and suppress background contributions, a five-constraint (5C) kinematic fit is performed by imposing energy-momentum conservation under the hypothesis of $\kk\pp \gamma\gamma\gamma$, with the invariant mass of two photons constrained to the known $\piz$ mass~\cite{pdg}.
%with an addition constraint on the $\gamma\gamma$ invariant mass to match the $\piz$ nominal mass.
The best combination is selected based on the minimum $\chi^{2}_{5\rm{C}}$, and events are required to satisfy $\chi^{2}_{5\rm{C}}<20$ for further analysis.
%By iterating through all possible photons combination, the combination of three photons, two of which are required to reconstruct a $\piz$, with a least $\chi_{5C}^{2}$ and satisfied with $\chi_{5C}^{2} < 20$ are chosen for further analysis.
This requirement is determined by optimizing the Figure-of-Merit $S/\sqrt{S+B}$, 
where $S$ is the number of signal events from the signal MC sample
and $S+B$ is the total number of events in data.
%where $S$ and $B$ are the numbers of signal and background events, respectively.
%The requirement is determined from MC simulations by optimizing $\bm{S/\sqrt{S+B}}$, where $\bm{S}$ and $\bm{B}$ are the numbers of signal and background events, respectively.
%The momenta of the charged tracks and the showers after the kinematic fit are used for further analysis throughout the text if not mentioned explicitly.

\subsection{Further selection}
\label{sub: fursel}

Additional selection criteria are applied to further suppress the significant background contributions. 
%A relatively wide mass window, $2800.0 <M_{\kk\ppp} < 3100.0~\mevcc$ is applied for $\etac$ signal.
To suppress the background from events without $\piz$ in the final state and to minimize the wrong combination of $\piz$ for the signal, the decay angle of $\piz$ is defined as
\begin{equation}
	\cos(\theta_{\rm{decay}})  = \frac{ E_{\gamma_1} - E_{\gamma_2}}{p_{\gamma_1 \gamma_2} },
\end{equation}
where $\gamma_{1}$ and $\gamma_{2}$ are the photons forming the $\piz$, 
$E_{\gamma_{1}}$ and $E_{\gamma_{2}}$ are the energies of photons, and $ p_{\gamma_1 \gamma_2}$ is the momentum of the two-photon system. A requirement of $|\cos(\theta_{\rm{decay}})|< 0.95$ is imposed. Studies based on the signal MC sample indicate that this $\piz$ decay angle requirement reduces the wrong combination rate of $\piz$ candidates to less than 1\%. 

\begin{figure}[htbp]
  \begin{center}
    \subfigure{\includegraphics[width=0.9\linewidth]{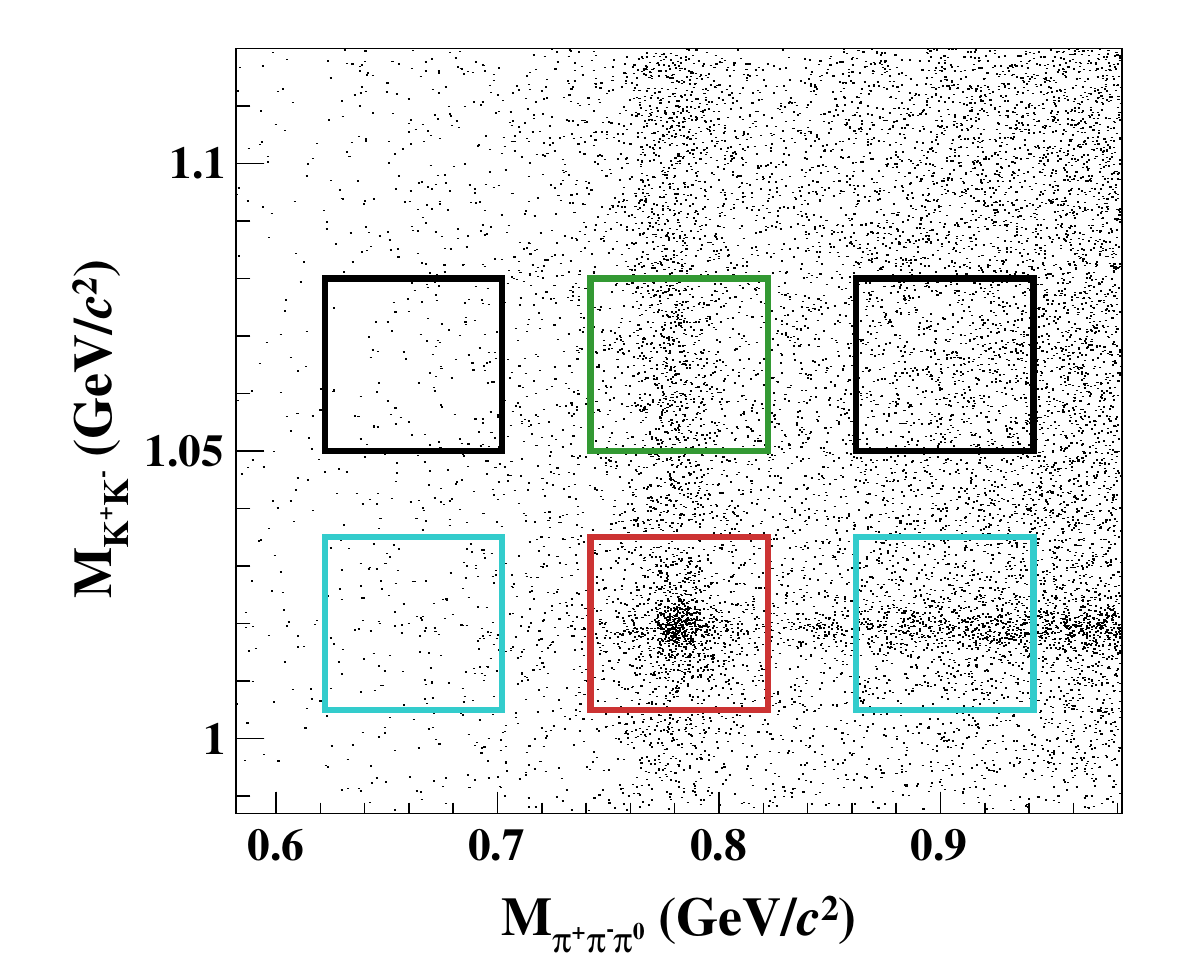}}
    \caption{The distribution of $\Mppp$ versus $\Mkk$.
			The horizontal and vertical bands correspond to the $\phi$ and $\omega$ signal regions, respectively.
			The red box represents the $\omega\phi$ signal region.
			The green box represents the $\phi$ sideband region (right $\omega$ and wrong $\phi$).
			The cyan boxes represent the $\omega$ sideband region (wrong $\omega$ and right $\phi$).
			The black boxes represent the non-$\omega / \phi$ region (wrong $\omega$ and wrong $\phi$).
			Because of the phase-space limitation there are no enough events below 0.9$\gevcc$ in $\Mkk$.
      }
    \label{fig:2Ddatashow}
  \end{center}
\end{figure}

After applying all the above selection criteria, the distribution of the invariant masses of $\ppp$ and $\kk$, 
$M_{\ppp}$ versus $M_{\kk}$ for the accepted candidates in data, is shown in Fig.~\ref{fig:2Ddatashow}. 
In this figure, a cluster of events is observed in the $\omega\phi$ signal region, along with a horizontal band for the $\phi$ signal and a vertical band for the $\omega$ signal.
The $\omega\phi$ signal region is defined as $| \Mppp - M_{\omega}^{\textnormal{PDG}} | <40.0\mevcc $,
and 
$| \Mkk - M_{\phi}^{\textnormal{PDG}} | <15.0\mevcc $, as shown in Fig.~\ref{fig:2Ddatashow}.
These signal regions are set to be 
approximately three times the fitted mass resolution around the known $\omega$ and $\phi$ masses~\cite{pdg}.
To further suppress backgrounds from processes such as  $\jpsi \to \phi\eta', \eta'\to\gamma \omega, \omega\to\ppp, \piz \to \gamma\gamma$, an $\eta'$ veto is applied. Events with an invariant mass $M_{\pip\pim 3\gamma}$ in the range $943.0<M_{\pip\pim 3\gamma}<969.0\mevcc$ are vetoed.
%The events within the mass window $943 < M_{\gamma\omega} < 969~\mevcc$ are vetoed to suppress the background from  $\jpsi \to \phi \eta^{\prime}, \eta^{\prime} \to \gamma \omega$.

\begin{figure}[!htp]
    \begin{center}
      \subfigure{\includegraphics[width=0.9\linewidth]{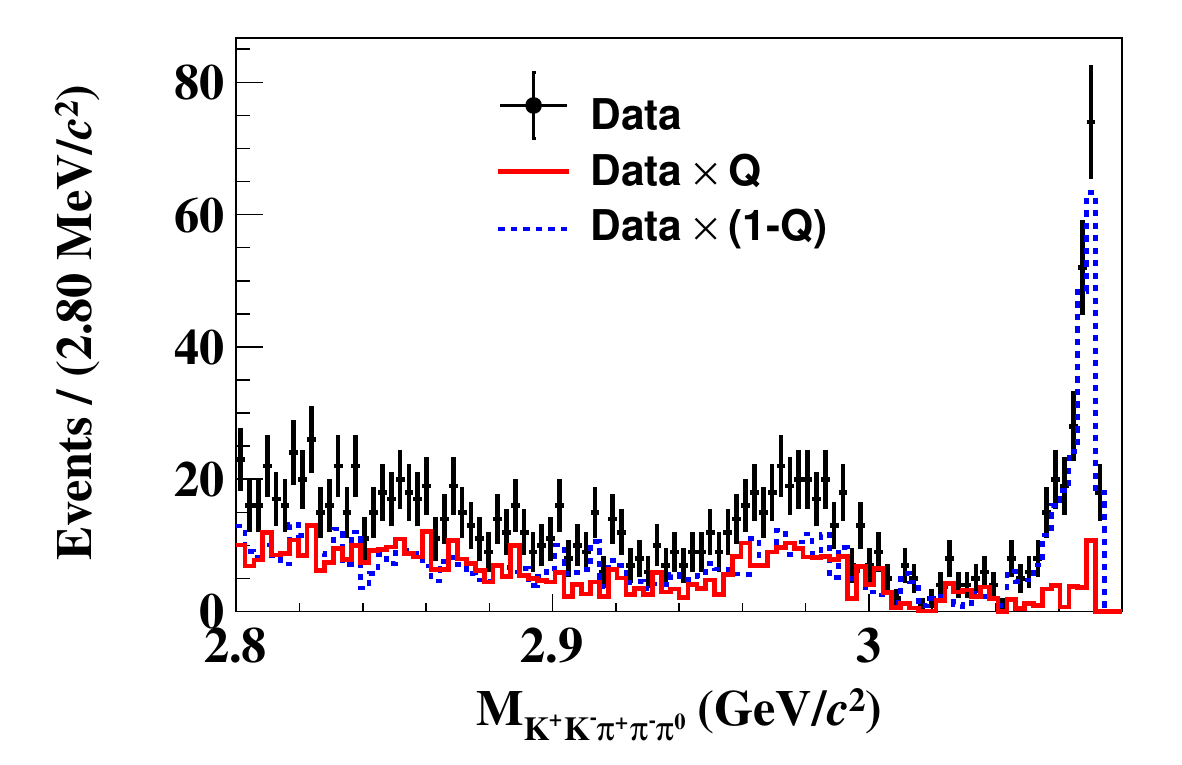}}
			\caption{The distribution of $M_{\kk\ppp}$ for the survived candidate events in the $\omega\phi$ signal region.
			The black dots represent data, the red line is for the Q-weighted events
			and the blue dotted line is for the (1 - Q)-weighted events.	
			These notations Q and 1 - Q represent the probabilities by which an event is defined to be signal and background, respectively (further detials will be in the next section).
			}
      \label{fig:DataWeightedmetac}
    \end{center}
\end{figure}

After applying all the event selection criteria, the distribution of the invariant mass $\Mkkppp$ for the surviving candidate events in data is shown in Fig.~\ref{fig:DataWeightedmetac}.
In this figure, the $\eta_c$ signal is clearly observed.
A prominent $\jpsi$ peak is also visible, indicating the presence of background events with a $\kk\ppp$ final state but without a radiative photon.
A more detailed discussion of this figure is provided in the second paragraph of Sec. IV. 

\section{BACKGROUND SUBTRACTION}
\label{BkgEstimation}

\subsection{Background study}
\label{sub_bkgstudy}
The potential background contributions are investigated based on the analysis of $\jpsi$ inclusive MC sample with a topology analysis tool~\cite{ZHOU2021107540}.
%For the signal decay $\etac\to\omega\phi$, 
The background is found to be relatively high and can be categorized into two types, i.e. background without $\omega$ or $\phi$ (namely non-$\omega / \phi$ background thereafter) and background with $\omega\phi$ but no $\etac$ intermediate state (namely non-$\etac$ background).
Due to possible decays such as $\etac\to\omega \kk$, $\etac\to\phi\ppp$ and $\etac\to\kk\ppp$ (without intermediate states), 
these backgrounds may produce a peak in the $\etac$ region of the $\Mkkppp$ distribution.
Because of the C-parity conservation, backgrounds from $\jpsi\to\omega\phi$ and $\jpsi\to\piz\omega\phi$ are forbidden.

\subsection{Non-$\omega\phi$ background estimation}
\label{sub_bkgestimation}

To estimate the contribution of the  non-$\omega / \phi$ background, a Q-weight method is employed.
This method, detailed in Ref.~\cite{Williams2009}, assigns a weight to each event, representing its probability of being signal or background. The Q-weight method has been widely used in experiments such as CLAS, Crystal Barrel and BESIII ~\cite{Williams2009a,Amsler2015,Ablikim2019}.
The sum of the Q-weights for all events corresponds to the number of signal events, while the sum of (1 - Q)-weights corresponds to the number of background events.
%The aim of Q-weight method is to obtain the signal weight for individual selected event candidate.
In the Q-weight method, a set of phase space (PHSP) variables $\vec{\xi}$ is used to select a control sample for each event. The PHSP distance $d_{i,j}^{2}$                between the $i$-th and $j$-th events is defined as: $d_{i,j}^{2}=\sum_{k} \left( \frac{\xi_{k}^{i}-\xi_{k}^{j}}{\Delta_k} \right)^2, $
where $\Delta_k$ is a normalization factor for the $k$-th variable. For each event, a control sample of 200 events with the smallest $d_{i,j}^{2}$             is selected.
  
Based on the selected control sample, a fit on the distribution of a set of discrimination variables $\vec{\nu}$ is carried out to extract the yields of signal and background, and the signal weight for the $i$-th event is obtained with the ratio of signal yield from the fit to the number of control sample.
The PHSP variables $\vec{\xi}$ and the corresponding normalization factors $\Delta_k$ are listed in Table~\ref{tab:Normalization}.
For events with $\Mkkppp < 2950.0\mevcc$, all PHSP variables are used. For events with $\Mkkppp \geq 2950.0\mevcc$, only the $m^2(\omega\phi)$ is used due to significant changes in its distribution.
The control sample is selected from candidate events within a broad range of $\Mkk$ and $\Mppp$: $\Mkk<1120.0\mevcc$ and $582.0<\Mppp<982.0\mevcc$ (see Fig.~\ref{fig:2Ddatashow}).  The variables ${\Mkk}$ and ${\Mppp}$are used as discrimination variables $\vec{\nu}$.
A two-dimensional (2-D) unbinned maximum likelihood fit is performed on the distribution of $\Mkk$ versus $\Mppp$ to extract the yields of $\omega\phi$ signal and non-$\omega / \phi$ background in the control sample.
In the 2-D fit, the $\phi$ signal in the $\Mkk$ distribution is modeled using the signal MC shape convolved with a Gaussian function, 
while the continuum background is described by an inverse-ARGUS function~\cite{Albrecht1990}. 
The $\omega$ signal in the $\Mppp$ distribution is modeled using the signal MC shape convolved with a Gaussian function, 
and the continuum background is described with a second-order Chebyshev polynominal.
In the fit, the signal MC sample is the process $\jpsi\to\gamma\omega\phi$ and with the similar PHSP variables $\vec{\xi}$ of data.
The overall probability density function (p.d.f.) in the 2-D fit is the sum of four components, 
$i.e.$ the $\phi$ and $\omega$ signal, the non-$\phi$ and $\omega$ background, the $\phi$ and non-$\omega$ background, 
as well as non-$\phi$ and non-$\omega$ background. Each component is the product of the corresponding functions for the $\Mkk$ and $\Mppp$ distributions.
With the above process, the Q-weights are obtained for the every individual candidate event with the signal yields from the above fit and the number of events of the control sample.
From the fit, the Q-weights for each candidate event are obtained as the ratio of the signal yield to the number of events in the control sample. Using these Q-weights, the distributions for Q-weighted signal events and (1-Q)-weighted non-$\omega / \phi$ background events are shown in Fig.~\ref{fig:DataWeightedmetac}. The Q-weight method for  estimating the non-$\omega / \phi$ background has been thoroughly validated using MC samples in this analysis.

\begin{table*}[htbp]%star makes the tab cross the two part
%\begin{table}[h!]
%\begin{center}
\centering
\caption{Definitions and normalization of coordinates in the multidimensional space.}
\begin{tabular}{c|c|c} \hline
%\toprule  %top wider line
%\hline\hline
Coordinate& Description& $\Delta_{k}$\\ \hline
%\midrule  %mid line
$\rm{cos}(\theta_{\gamma})$& Polar angle of the radiative photon& 2\\
$\rm{cos}(\theta_{\omega})$& Polar angle of the $\omega$ & 2\\
$\varphi_{\omega}$& Azimuthal angle of the $\omega$ in $\eta_{c}$ rest frame& 2$\pi$\\
$\rm{cos}(\theta_{\phi})$& Polar angle of the $\phi$& 2\\
$\varphi_{\phi}$& Azimuthal angle of the $\phi$ in $\eta_{c}$ rest frame & 2$\pi$\\
$\rm{cos}(\theta_{K^{+}})$& Polar angle of the $K^{+}$ & 2\\
$\varphi_{K^{+}}$& Azimuthal angle of the $K^{+}$ & 2$\pi$\\
$m^{2}(\omega\phi)$& Invariant mass square of the $\omega\phi$ system& Standard deviation\\
$m^{2}(\gamma\omega)$& Invariant mass square of the $\gamma\omega$ system& Standard deviation\\
$m^{2}(\gamma\phi)$& Invariant mass square of the $\gamma\phi$ system& Standard deviation\\
$\lambda_{\omega}/\lambda_{\rm{max}}$& Normalized slope parameter of the $\omega$ Dalitz plot& 1\\ \hline
%\bottomrule %bottom wider line
\end{tabular}
\label{tab:Normalization}
%\end{center}
\end{table*}

\section{Signal extraction}
\label{DATAANA}

\subsection{Fitting method}

%As discussed previously, the surviving candidate events are shown in Fig.~\ref{fig:DataWeightedmetac}, including the background with $\omega\phi$ and non-$\omega\phi$ components, and $\etac\to\omega\phi$ signal is evidenced after the application of Q-weight method.
In Fig.~\ref{fig:DataWeightedmetac}, the $\etac\to\omega\phi$ signal is evident after adopting the Q-weight method.
To extract the $\etac$ signal yield, an unbinned maximum likelihood fit is performed on the $\Mkkppp$ distribution.
%  is carried out with $\Mkkppp>2.8~\gevcc$.
%In the fit, three components, the signal $\etac\to\omega\phi$, the backgrounds with $\omega\phi$ and non-$\omega\phi$ structure, are considered, and the overall p.d.f is the sum of above three components.
%{\scriptsize
%{\small
%\begin{equation}
%	\begin{aligned}
%	{f_{all}(\Mkkppp)}
%	=
%	{\rm Shape}_{M_{\eta_{c}}}^{\textnormal{MC}} + {\rm Shape}_{\textnormal{Q-weight}} + \textnormal{Argus}.
	%{\rm Shape}_{M_{\eta_{c}}}^{{\rm MC}} + {\rm Shape}_{{\rm Q-weight}} + Argus
	%{\rm Shape}_{m_{\eta_{c}}}^{{\rm MC}} + 3^{rd} {\rm Cheby} + {\rm Shape}_{{\rm Q-weight}}
%	\label{eq:PDFequation}
%	\end{aligned}
%\end{equation}
%}
%%%%%%%%%%%%%%%%%%%%%%%%%%%%%%%%%%%%%%%%%%%%%%%%%%%%%%%%%%%%%%%%%%%%%%%%%%%%%%%%%%%%%%%%%%%%%%%%%%%%
In the fit, 
%the p.d.f of signal decay $\etac\to\omega\phi$ is extracted with the signal MC sample.
%The line shape of $\etac$ produced in $\jpsi$ M1 transition is given by:
the line shape of the $\eta_{c} \to \omega \phi$ signal is modeled using the MC simulated sample generated by an  M1 transition: 
\begin{equation}
%\begin{split}
E_{\gamma}^{3} \times BW(m) \times f_{d}(E_{\gamma}),
%	(E_{\gamma}^{3} \times BW(m) \times f_{d}(E_{\gamma}) \times f_{\epsilon}(m) \times f_{\textnormal{PHSP}}(m) ) %\\
%\otimes G(\delta m, \sigma),
			\label{eq:lineshapeetac}
%\end{split}
\end{equation}
		%\begin{equation}
			%(E_{\gamma}^{3} \times BW(m) \times f_{d}(E_{\gamma})\times f_{\epsilon}(m) \times f_{PHSP}(m) ) \otimes G(\delta m, \sigma)
			%\label{eq:lineshapeetac}
		%\end{equation}
		where $E_{\gamma}$ is the energy of the radiative photon as described in  Sec.~\ref{sec:detmc}, and $BW(m)$ is the Breit-Wigner function of the $\etac$,
		\begin{equation}
			BW(m)= \left \vert \frac{1}{m^{2}-M_{R}^{2}+i  M_{R}  \Gamma_{R}} \right \vert ^{2} ,
			%BW(m)= \frac{1}{m^{2}-M_{R}^{2}+i \times M_{R} \times \Gamma_{R}}
			\label{eq:RelaBW}
		\end{equation}
		where $m=M_{K^+K^-\pi^+\pi^-\pi^0}$,
		and $f_{d}(E_{\gamma})$ is a function that damps the diverging tail arising from the $E_{\gamma}^{3}$ dependence.
		%the $f_{\epsilon}(m)$ is the efficiency;
		%the $f_{PHSP}(m)$ is the PHSP factor
		%and the $G(\delta m, \sigma)$ is a Gaussian function describing the mass shift and detector resolution.
In this analysis, the damping function $f_{d}(E_{\gamma})$~\cite{Brambilla2006} adopted from the CLEO collaboration~\cite{Mitchell2009} and is given by:
		\begin{equation}
			f_{d}(E_{\gamma})={\rm exp}(-E^{2}_{\gamma}/8 \beta^{2}),
		\end{equation}
where the form factor parameter $\beta = 65.0\pm2.5\mev$. 
%		Another suitable function developed by CLEO considering the overlap of wave functions is demonstrated as~\cite{Mitchell2009,Brambilla2006}:
%		 from CLEO fit result.
%		Both two functions are reasonable in this analysis, so we choose the CLEO result for our $\eta_{c}$ line shape.

%%%%%%%%%%%%%%%%%%%%%%%%%%%%%%%%%%%%%%%%%%%%%%%%%%%%%%%%%%%%%%%%%%%%%%%%%%%%%%%%%%%%%%%%%%%%%%%%%%%%

%	\item Shape of Q-weight estimated backgrounds :
The non-$\omega / \phi$ background is described by the p.d.f. extracted from the corresponding background curve in Fig.~\ref{fig:DataWeightedmetac}, with its amplitude fixed to the results obtained from the Q-weight method.
%The p.d.f of non-$\omega\phi$ background is extracted from the  corresponding background curve in Fig.~\ref{fig:DataWeightedmetac}, and its amplitude is fixed to the results from the Q-weight method.
%		As mentioned previously, all the backgrounds without ($\omega - \phi$) structure will be estimated by Q-weight method.
%		This kind of data-based method can well confront the circumstance that the MC sample can not accurately simulate the real data.
%		In our fit, firstly the estimated background will be obtained by weighting the data, which is the blue line in Figure~\ref{fig:DataWeightedmetac}.
%		Then a corresponding PDF is generated based on these event and included into the fit.
%		To be noticed that the event number corresponding to the generated PDF in the fit is constrained to the estimated background event number.
%	\item Shape of non-peaking backgrounds :
The background from $\jpsi\to\gamma \omega\phi$ without the $\etac$ signal is described by an ARGUS function~\cite{Albrecht1990}, with its parameters left free in the fit.
Previous studies indicate no significant structures in the distribution within the region of interest.
Additionally, contributions from resonances peaking in the low-mass region are expected to be small, based on the results of amplitude analyses~\cite{Ablikim2006,Ablikim2013a}.
Interference between the $\etac$ signal and the continuum $0^{-+}$ contribution is not considered in the fit.

%		In this analysis, the non-peaking backgrounds in $M_{\pi^{+}\pi^{-}\pi^{0}K^{+}K^{-}}$ associated with the $\omega-\phi$ structure are described with this term.
%		The backgrounds are $J/\psi \rightarrow \gamma\omega\phi$ and $J/\psi \rightarrow \gamma X \rightarrow \gamma\omega\phi$ decay modes.
%		The $J/\psi \rightarrow \gamma\omega\phi$ PHSP decay mode could be well simulated by exclusive MC sample, d is verified to be well modeled by an Argus Function~\cite{Albrecht1990} in our signal region.
%For the $J/\psi \rightarrow \gamma X \rightarrow \gamma\omega\phi$ decay mode, no obvious significant structure is observed in the $M_{\pi^{+}\pi^{-}\pi^{0}K^{+}K^{-}}$ spectrum above 2.8 GeV/$c^{2}$.
%		Additionally, based on the Partial Wave Analysis (PWA) result for $J/\psi \to \gamma\omega\phi$ ~\cite{Ablikim2006,Ablikim2013a}, there are no significant structures near 2.8 GeV/$c^{2}$.
%		Notably, even the closest potential structure is $\eta(2500)$,
%		so those structures below 2.8 GeV/$c^{2}$ also behave the same as the $J/\psi \rightarrow \gamma\omega\phi$ PHSP decay mode in our signal region,
%		and could also be described by an Argus Function.
%		Therefore, the Argus Function is adopted to model all these backgrounds with the $\omega-\phi$ structure.

%\end{itemize}

\begin{figure}[!htp]
    \begin{center}
      \subfigure{\includegraphics[width=0.9\linewidth]{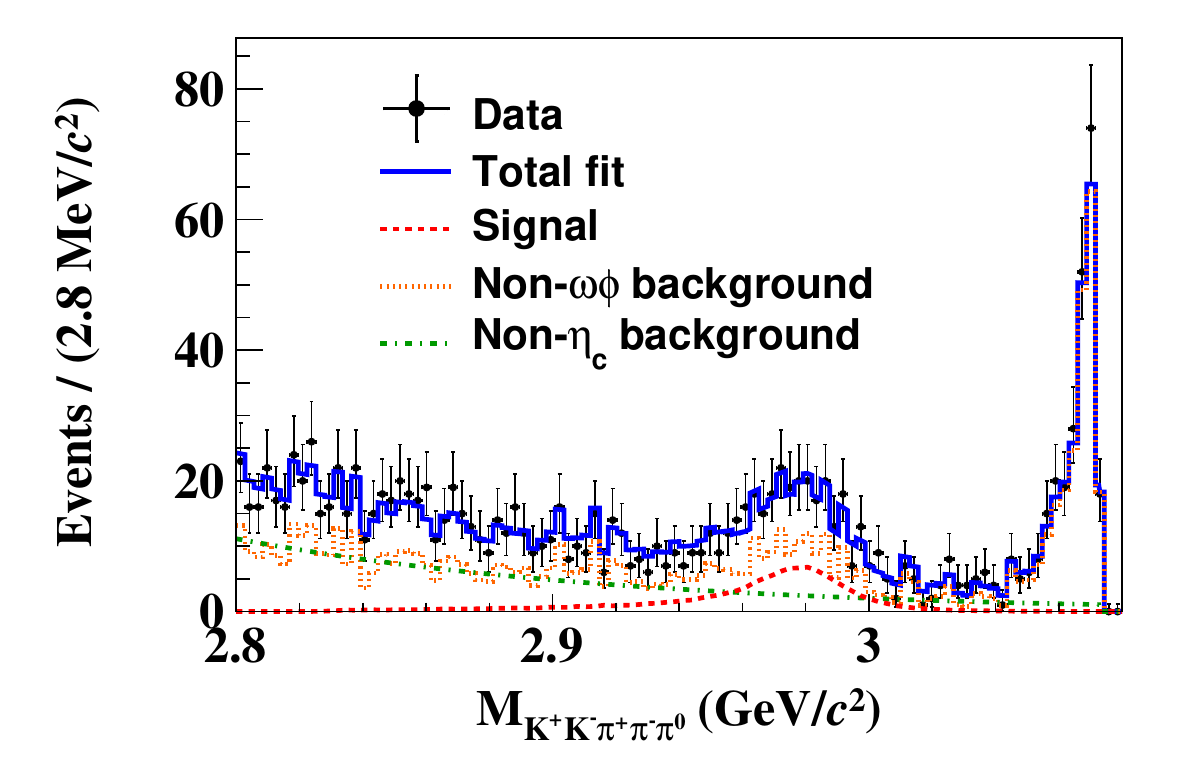}}
			\caption{The $M_{\kk\ppp}$  distribution of the accepted candidates for $\eta_c\to \omega\phi$.
                The black dots with error bars represent data and the blue curve represents the accumulated fit curve.
				The red dashed curve represents the signal,
				the orange dotted curve and green dash-dot curve represent the non-$\omega / \phi$ background estimated by Q-weight method and the non-$\etac$ background, respectively.}
			\label{fig:metacfitresultdata}
    \end{center}
\end{figure}

The fitting result is shown  in Fig.~\ref{fig:metacfitresultdata}.
The obtained $\etac$ signal yield is $N_{\rm{obs}} = 118\pm 28$, where the uncertainty is statistical only. 
An input-output check is performed to validate the reliability of the fitting procedure, and no significant bias is observed.
%

%%%%%%%%%%%%%%%%%%%%%%%%%%%%%%%%%%%%%%%%%%%%%%%%%%%%%%%%%%%%%%%%%%%%%%%%%%%%%%%%%%
\subsection{Numerical result}

The branching fraction of $\etac \rightarrow \omega\phi$ is calculated using
\begin{equation}
%\begin{equation}
	\BR(\etac \rightarrow \omega\phi) = \frac{N_{\rm{obs}}}
{ N_{\jpsi} \cdot \BR_{\jpsi} \cdot \BR_\phi \cdot \BR_\omega \cdot \BR_{\piz}  \cdot \epsilon}, 
	\label{eq:Brcalculation}
\end{equation}
where $N_{\jpsi}$ is the total number of $\jpsi$ events, and $\epsilon=5.01\%$ is the signal detection efficiency. 
$\BR_{\jpsi}$, $\BR_\phi$, $\BR_\omega$ and $\BR_{\piz}$ are the branching fractions of the decays 
$\jpsi\to\gamma \etac$, $\phi\to\kk$, $\omega\to\ppp$ and $\pi^0\to\gamma\gamma$, 
respectively, cited from the PDG~\cite{pdg}. 
With these inputs, the branching fraction is determined to be $\BR(\etac\rightarrow\omega\phi) = (3.86 \pm 0.92) \times 10^{-5}$, 
where the uncertainty is statistical only. This result is consistent with the corresponding upper limit of $2.5 \times 10^{-4}$ at the 90\% confidence level reported in the PDG.

% by Eq.~\ref{eq:Brcalculation},
%where $N_{total}$ is the number of $J/\psi $ sample which is $1.009 \times 10^{10}$,
%and the $\epsilon$ is the efficiency calculated from signal MC sample which is found to be 5.05\%.
%The PDG existing upper limit for $Br(\eta_{c} \rightarrow \omega\phi)$ is 
%Eventually the $Br(\eta_{c} \rightarrow \omega\phi)$ in this analysis is calculated as $$.
%and the signal significance is calculated as 4.6$\sigma$.

\section{Systematic uncertainties}
\label{sec:sys}

Several sources of systematic uncertainty are considered in the measurement of the branching fraction, including external input parameters, signal detection efficiency, and signal extraction.

%the sources of systematic uncertainties in the Branching Fraction measurement of $\eta_{c}\rightarrow\omega\phi$.
%The total systematic uncertainty is 26.7\%.
%and the details are introduced as following individually.
%The systematic uncertainty from $J/\psi \rightarrow \gamma\eta_{c}$ decay is a dominant contribution, which is 23.6\%,
%and the values in bracket are obtained by excluding this out.

%%%%%%%%%%%%%%%%%%%%%%%%%%%%%%%%%%%%%%%%%%%%%%%%%%%%%%%%%%%%%%%%%%%%%%%%%%%%%%%%%%%%%%%%%%%%%%%%%%%%

		The uncertainty of the total number of $J/\psi$ events is 0.5\% according to Ref.~\cite{Ablikim2022}.
		The uncertainties in the branching fractions of  $\jpsi \rightarrow \gamma\etac$, $\omega \rightarrow \ppp$.
		$\phi \rightarrow \kk$ and $\piz\rightarrow\gamma\gamma$, taken from the PDG~\cite{pdg}, is 10.0\%.

		The tracking efficiencies for charged pions and kaons are studied with the control samples of 
		$\psi' \rightarrow\pp\jpsi \rightarrow \pp l^{+}l^{-}$ and $\jpsi \rightarrow \rho^{0}\piz\rightarrow \ppp$~\cite{Ablikim2011}, 
	as well as $\jpsi \rightarrow K^{0}_{S}\bar{K}^{*0}(892)+c.c. \rightarrow K^{0}_{S}K^{+}\pi^{-}+c.c. $~\cite{Ablikim2011}, respectively.
    The efficiency difference between data and MC simulation is less than 1\% per pion or kaon.
		Therefore, the total systematic uncertainty due to tracking for the four charged particles is assigned as 4.0\% in  total.
		%Two pions and two kaons are identified in this analysis.
		The PID efficiencies for charged pions and kaons are estimated with  control samples of 
		$\jpsi \rightarrow \ppp$~\cite{Ablikim2011} and $\jpsi \rightarrow \kk\piz$~\cite{Ablikim2011}, respectively.
		The efficiency difference between data and MC simulation is less than 1\% per pion or kaon, leading to a total systematic uncertainty of 4.0\% for PID.
The photon detection efficiency is studied with the control sample of
$\jpsi \rightarrow \rho^{0}\piz\rightarrow \ppp$~\cite{Ablikim2010}.
The difference in the efficiency between data and MC simulation, 1\%, is taken as the uncertainty of per photon, 
%The systematic uncertainty for photon detection is estimated to be 1\% per photon.
resulting in a total systematic uncertainty of 3.0\% for the three photons.
		The uncertainty in the kinematic fit arises primarily from inconsistencies in the kinematic variables of charged tracks. A helix parameter correction is applied to the signal MC sample to minimize data-MC differences~\cite{Ablikim2013}.
		%The detail of the corresction of interesting disrtibutions are given in Appendix.~\ref{app:kmfun}.
		%The correction parameters for charged tracks are refers to Ref.~\cite{Ablikim2013}.
    The resulting change in detection efficiency, 6.0\%, is considered as the systematic uncertainty.
		The uncertainties of $\phi$ and $\omega$ signal regions are due to resolution differences between data and MC simulation. 
		Fits to the $\Mppp$ and $\Mkk$ distributions of data are performed using the corresponding MC simulation shapes convolved with Gaussian functions. 
The resulting changes in detection efficiency, 0.5\% for $\omega$ and 0.1\% for $\phi$, are assigned as systematic uncertainties.
		
    The uncertainty of the $\piz$ decay angle requirement is studied with the control sample of $\jpsi \to \ppp$.
    The efficiency difference between data and MC simulation, 1.0\%, is taken as the systematic uncertainty.
    The uncertainty of the $\eta^{\prime}$ veto is studied using an alternative signal MC sample generated by varying the input form factor parameter  $\beta$ by $\pm 1 \sigma$.
    The resulting efficiency difference, 0.1\%, is taken as the systematic uncertainty.

	%\item Signal MC shape :
		
     The uncertainty of the $\etac$ signal MC shape is studied with alternative signal MC samples.  These include variations in the signal model parameters by $\pm 1 \sigma$ and the use of a damping factor employed by the KEDR experiment~\cite{Anashin2011,Anashin2014}.
		The alternative option of $f_{d}(E_{\gamma})$ used by the KEDR experiment~\cite{Anashin2011,Anashin2014}:
%		There are two reasonable options of the damping function $f_{d}(E_{\gamma})$.
%		The first suitable function used by KEDR is demonstrated as ~\cite{Anashin2011,Anashin2014}:
		\begin{equation}
			%\dfrac{f_{d}(E_{\gamma})=\frac{E^{2}_{0}}{E_{0}E_{\gamma}+(E_{\gamma}-E_{0})^{2}}}
			f_{d}(E_{\gamma})=\frac{E^{2}_{0}}{E_{0}E_{\gamma}+(E_{\gamma}-E_{0})^{2}},
			\label{eq:lineshapeetacKEDR}
		\end{equation}
		where $E_{0}=\frac{m_{\jpsi}^{2}-m_{\etac}^{2}}{2m_{\jpsi}}$.
    %The analyses are repeated with different alternative signal MC sample, individually, 
   The largest change in the measured branching fraction, 1.5\%, is taken as the systematic uncertainty.
    The uncertainty due to the $\eta_{c}$ width  is studied by varying the world average $\eta_{c}$ width within its uncertainty. The resulting largest change in the branching fraction, 3.0\%, is assigned as the systematic uncertainty.
	 	The uncertainty of the shape of the $J/\psi\to\gamma\omega\phi$ non-peaking background is studied with an alternative fit performed by replacing the ARGUS function with a third-order Chebyshev polynomial. 
		The resulting change in the signal yield, 2.5\%, is taken as the corresponding uncertainty.
		The uncertainty of the Q-weight method is studied with MC sample.
		MC sample containing background and signal are generated and subjected to Q-weight method.
		The deviation between the number of generated signal events and the sum of obtained Q-weight is determined to be 7.6\%, which is taken as the systematic uncertainty of the method.
	%\item Fit range:
		The uncertainty in the fit range is studied by performing alternative fits with different ranges: $[2775.0 , 3100.0] \mevcc$ and $[2825.0 , 3100.0] \mevcc$. The largest change in the signal yield, 1.0\%, is taken as the systematic uncertainty.

%The statistical significance changes throughout the systematic is also checked, and the minimum value is determined as the final signal significance, which is found to be 4.0$\sigma$.
%The minimum value of significance occurs in the systematic check for the background estimation and fitting method.
%The corresponding value is found to be 4.0$\sigma$ and hence is regarded as the final significance result.

\begin{table}[htbp]
	%\begin{center}
	\centering
	\caption{Sources and assigned systematic uncertainties in the branching fraction measurement.}
	%\resizebox{\textwidth}{!}{%auto resize text size
		\begin{tabular}{cc} \hline
			%\toprule  %top wider line 
				Source & Uncertainty~(\%)\\ \hline
			Total number of ${J/\psi}$ events & 0.5\\ 
			Intermediate branching fractions & 10.0 \\
			Tracking efficiency& 4.0\\
      PID & 4.0\\
			Photon detection & 3.0\\
			Kinematic fit & 6.0\\
			$\omega$ signal region & 0.5\\
			$\phi$ signal region & 0.1\\
			Decay angle of ${\pi^{0}}$ & 1.0\\
			${\eta'}$ veto & 0.1\\
			Signal MC shape of $\eta_{c}$ & 1.5\\
			$\eta_{c}$ width & 3.0\\
			Non-peaking background description & 2.5\\
			Non-$\omega\phi$ background & 7.6 \\
			Fit range & 1.0\\ \hline 
			%\midrule  %mid line
			Total & 16.0\\ \hline 
			
			%\bottomrule %bottom wider line
		\end{tabular}
		%}
	\label{tab:uncertainties}
	%\end{center}
\end{table}

All systematic uncertainties are summarized in Table~\ref{tab:uncertainties}. 
The total systematic uncertainty is calculated as the square root of the quadratic sum of the individual contributions.
The $\etac$ signal significance is estimated from the change of likelihood and the change of number of degrees of freedom with and without the $\etac$ signal included in the fit.
The significance varies during systematic checks, and the minimum value, 4.0$\sigma$, is taken as the final signal significance.

%The overall systematic uncertainty is 26.7\%, been dominant by that of branching fraction $\jpsi\to\gamma \etac$, 23.6\%.

\section{SUMMARY AND DISCUSSION}
\label{sum}
Using a sample of $(10087\pm 44) \times 10^6$ $\jpsi$ events~\cite{Ablikim2022} collected with the BESIII detector at BEPCII,
we report the first evidence for the doubly OZI-suppressed decay $\eta_{c} \to \omega\phi$ with a significance of 4.0$\sigma$.
This result is achieved after effectively suppressing the high and complex backgrounds using the Q-weight method. 
Although potential interference effects are not considered in this analysis, they may have an impact on the measured branching fraction.
The branching fraction of $\etac \to \omega\phi$ is determined  to be $\BR(\etac \to \omega\phi) = (3.86 \pm 0.92 \pm 0.62) \times 10^{-5}$,
where the first uncertainty is statistical and
the second is systematic.
%The systematic uncertainty is dominant because of the high uncertainty of decay branching fraction of $\jpsi \to \gamma \etac$.
This result, together with the previously measured branching fractions of $\etac \to \omega\omega$ and $\etac \to \phi\phi$~\cite{Ablikim2019,Ablikim2017a}, provides robust inputs for studying the mechanisms of OZI-suppressed decays in $\jpsi$ and $\eta_c$ systems~\cite{Seiden1988}~\cite{Haber1985}. Furthermore, this analysis offers important insights for establishing theoretical models based on the Helicity Selection Rule (HSR) and contributes to a deeper understanding of the  $\etac \to VV$ process.

\begin{acknowledgments}
The BESIII collaboration thanks the staff of BEPCII, the IHEP computing center and the supercomputing center of the University of Science and Technology of China (USTC) for their strong support.
This work is supported in part by National Key R\&D Program of China under Contracts Nos. 2020YFA0406300, 2020YFA0406400, 2023YFA1609400, 2023YFA1606000; 
National Natural Science Foundation of China (NSFC) under Contracts Nos. 11625523, 11635010, 11735014, 11935015, 11935016, 11935018, 12025502, 12035009, 12035013, 12061131003, 12105276, 12122509, 12192260, 12192261, 12192262, 12192263, 12192264, 12192265, 12221005, 12225509, 12235017, 12361141819; 
the Chinese Academy of Sciences (CAS) Large-Scale Scientific Facility Program; 
the CAS Center for Excellence in Particle Physics (CCEPP); 
Joint Large-Scale Scientific Facility Funds of the NSFC and CAS under Contracts Nos. U2032111, U1732263, U1832103, U1832207; 
CAS Youth Team Program under Contract No. YSBR-101;
CAS under Contract No. YSBR-101; 
100 Talents Program of CAS; 
The Institute of Nuclear and Particle Physics (INPAC) and Shanghai Key Laboratory for Particle Physics and Cosmology; 
Agencia Nacional de Investigación y Desarrollo de Chile (ANID), Chile under Contract No. ANID PIA/APOYO AFB230003; 
German Research Foundation DFG under Contract No. FOR5327; 
Istituto Nazionale di Fisica Nucleare, Italy; 
Knut and Alice Wallenberg Foundation under Contracts Nos. 2021.0174, 2021.0299; 
Ministry of Development of Turkey under Contract No. DPT2006K-120470; 
National Research Foundation of Korea under Contract No. NRF-2022R1A2C1092335; 
National Science and Technology fund of Mongolia; 
National Science Research and Innovation Fund (NSRF) via the Program Management Unit for Human Resources \& Institutional Development, Research and Innovation of Thailand under Contract No. B50G670107; 
Polish National Science Centre under Contract No. 2019/35/O/ST2/02907; 
Swedish Research Council under Contract No. 2019.04595; 
The Swedish Foundation for International Cooperation in Research and Higher Education under Contract No. CH2018-7756; 
U. S. Department of Energy under Contract No. DE-FG02-05ER41374

\end{acknowledgments}
	
\bibliography{main}
	
\end{document}